\documentclass[a4paper, runningheads]{llncs}
\usepackage{makeidx}  % allows for indexgeneration
\usepackage{listings}

\usepackage{amsthm}
\theoremstyle{definition}

\usepackage[table]{xcolor}
\usepackage{amsmath}
\usepackage{setspace}
\usepackage{caption}
\usepackage{csquotes}
\usepackage{booktabs}
\usepackage{tabularx}
\usepackage{listings}
\usepackage{comment}
\usepackage{enumerate}
\usepackage{float}
\usepackage{xcolor}
\usepackage{nameref}
\usepackage{hyperref}
\usepackage{multirow}
\usepackage{makecell}
\usepackage{diagbox}
\usepackage{algorithm2e}
\RestyleAlgo{ruled}
\usepackage{longtable}
\usepackage{enumitem}
\usepackage{graphicx}
\usepackage{wrapfig}
\usepackage{todonotes}
\usepackage{amssymb}   % basic math symbols
\usepackage{pifont}

\newcommand\nk[1]{\textcolor{black}{#1}}

\newcommand\etm[1]{\textcolor{black}{#1}}

\newcommand{\orcidJM}	{\href{https://orcid.org/0000-0002-6332-5801}{\protect\includegraphics[scale=0.045]{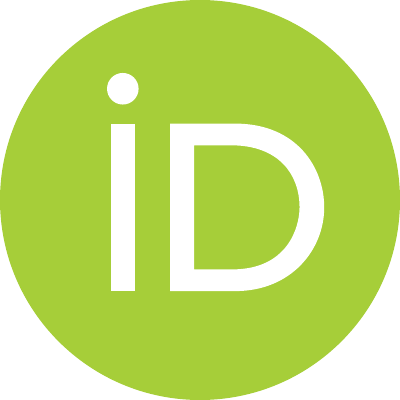}}}
\newcommand{\orcidSRM}	{\href{https://orcid.org/0000-0001-5656-6108}
{\protect\includegraphics[scale=0.045]{orcid}}}
\newcommand{\orcidNK}	{\href{https://orcid.org/0009-0009-9010-2855}
{\protect\includegraphics[scale=0.045]{orcid}}}

\begin{document}

\author{Nataliia Klievtsova\inst{1} 
\orcidNK{}
\and Juergen Mangler\inst{1}
\orcidJM
\and Stefanie Rinderle-Ma\inst{1}
\orcidSRM
}
\institute{$^1$Technical University of Munich, Garching, Germany\\%
TUM School of Computation, Information, and Technology\\%
\email{{firstname.lastname}@tum.de}
}

\mainmatter % start of the contribution
\title{Digital Innovation through Knowledge Processes}
\titlerunning{Digital Innovation through Knowledge Processes} 

\maketitle
\begin{abstract}
The artefact at the intersection of knowledge and process management is the process, which describes how enterprises are generating value. In knowledge management literature the relation of knowledge and processes is discussed, often leading to the definition of knowledge intensive processes, which entail a high level of human involvement. The process management community on the other hand focuses on the formalization of processes as models that can be executed, monitored, and improved. In this paper we explore the relationship of process resources such as data, objects, artefacts and humans, in order to come to more universal definition of knowledge intensive processes. For this purpose, we analyze different patterns that frequently occur in knowledge management processes, how they are represented in process models, and which types of knowledge are represented by them. The results comprise a categorization of process models into 6 categories that allow to easily see the knowledge intensity as well as a collection of common knowledge process patterns. The patterns were derived from a real-world knowledge gathering process, which includes a wide array of patterns and concepts detailed in this paper. We think that a holistic view on knowledge intensive processes, how to model, execute, monitor, and asses their impact, will speed up and improve the quality of digital transformation projects.
%The abstract should briefly summarize the contents of the paper in 150--250 words.
\end{abstract}


% \maketitle

\section{Introduction}
\label{sec:intro}

While knowledge has always been crucial, the shift from an industrial society based on the manufacturing and assembling of goods to a post-industrial society, where the service sector became predominant, makes knowledge the key economic resource~\cite{38db798c34c640d8804e7f955f1ce9f3}. Thus, the effective management of knowledge and information becomes a critical issue for organizations in the modern economy to achieve a competitive advantage~\cite{doi:10.2307/41165944}.
Knowledge management (KM) provides a competitive sustainable advantage for companies and organizations \cite{MAHDI2019320}.
According to Dalkir \textsl{``[k]nowledge management is the deliberate and systematic coordination of an organization’s people, technology, processes, and organizational structure in order to add value through reuse and innovation. This is achieved through the promotion of creating, sharing, and applying knowledge as well as through the feeding of valuable lessons learned and best practices into corporate memory [...]''}~\cite{dalkir2013knowledge}.

At the intersection of KM and Business Process Management (BPM), especially when considering knowledge management processes, several advantages are to be gained, as sticking to a formal process structure and a definition which activities are performed by whom provides clarity, and also helps to improve process performance~\cite{brocke2010handbook} as well as business performance~\cite{Jordan}. Especially process-oriented KM offers task-related knowledge to employees within the organization’s operative business processes in a more targeted way~\cite{process-knowledge}, avoiding their information overload and helping them to concentrate only on important information. In addition, KM in general and KM processes in specific play an essential role in facilitating the implementation of many information systems~\cite{al-emran_impact_2018}.
. 

Due to the inherent characteristics of process models, integrating human users, systems, physical devices, and services, and the related data and knowledge, process modeling enables knowledge transformation from its informal into formal representation facilitating KM processes~\cite{DBLP:journals/jkm/KalpicB06} (e.g., knowledge acquisition, sharing, application, protection, storage, creation, and IS knowledge~\cite{al-emran_impact_2018}).
The widespread adoption of BPM methodologies and technologies, along with growing interest in process-oriented KM, has led to the exploration of new types of processes, commonly referred to as Knowledge-Intensive Processes~\cite{DBLP:journals/jodsn/CiccioM015,DBLP:conf/hicss/MarjanovicF11}.
Handling knowledge-intensive processes motivates the development of new paradigms that shift from activity-centric approaches to data-centric methods (e.g., data-aware processes \cite{DBLP:journals/smr/KunzleR11}, artefact-centric processes \cite{DBLP:journals/fgcs/OriolGET21}, product-based processes~\cite{a9e7410a34844143a44a58bbdaf2b670}), which focus primarily on data objects within the process. However, none of these methods has yet been as widely accepted, supported, or adopted as traditional activity-centric approaches~\cite{DBLP:conf/modellierung/Breitmayer0PR24}.

\textbf{Problem Statement:} Despite the high variety of KM processes, offering a fertile ground for process technology integration, and potential benefits achieved by KM-BPM integration, the usage of process technology (i.e., process modeling, analysis, and automation) for KM processes has not been exploited in full extend yet. Worse yet, even the definition what knowledge(-intensive) processes are is still fuzzy and contested (see Sec. \ref{sec:relwork} and Sec. \ref{sec:kps}). 
\nk{In addition, with the advent of LLMs, genAI tools and autonomous agents are now actively participating in knowledge generation within the process, transforming them from passive resources into active knowledge co-constructor~\cite{10.1108/JKM-02-2025-0252}, blurring the traditional distinction between human-centered knowledge activities and system-supported knowledge processing. } This leads to the following research questions tackled in this work:

\noindent\textbf{RQ1:} To what extent are knowledge(-intensive) processes necessarily human-centered (as related work implies), and how to distinguish whether a business process should be considered knowledge-intensive?

\noindent\textbf{RQ2:} Which patterns characterize knowledge processes when utilizing a traditional activity-centric approach, where knowledge considered as process/task input and output?

The remainder of this paper is structured as follows: Section \ref{sec:relwork} 
discusses related work. Section \ref{sec:kps} tackles RQ1 and provides a definition for knowledge processes as well as a
process characterization method, that allows to precisely identify different types of knowledge processes, and their knowledge intensity. 
Section \ref{sec:patterns} elaborates RQ2, based on a real-world process resulting in a (non-exhaustive) list of $31$ patterns.
We conclude the paper in Sec. \ref{sec:conc}.

\section{Terminology and Related Work}
\label{sec:relwork}

According to~\cite{DBLP:journals/jkm/BakerBTD97}, \textit{knowledge} can be considered an asset that is created when information, combined with experience, skills, and personal capabilities, is applied to the content and, more importantly, the context of a problem. A \textit{process} is a series of activities that interact to produce a result. In the context of \textit{business}, a \textit{process} refers to a structured set of activities within an enterprise, describing their logical order and dependencies, with the objective of producing a desired outcome~\cite{AGUILARSAVEN2004129}. 
Knowledge and processes are closely connected. On one hand, most organizational knowledge is collected during process execution. On the other hand, knowledge is required to execute these processes and improve them~\cite{yousefiyan2009combination}. However, without proper management, both knowledge and processes may remain inefficient.

Since Knowledge Process Management (KPM) and Business Process Management (BPM) are interdependent, they have been brought together in the management context. Organizations can combine the benefits of both BPM and KPM to improve workflow efficiency~\cite{PASCHEK2018182}, facilitate knowledge sharing, optimize resource utilization, and enhance collaboration. Implementing Process-Oriented Knowledge Management enables this integration by embedding KM activities into business processes and defining interfaces to the corresponding Knowledge Processes ~\cite{remus2003blueprint}.

\begin{figure}[htb]
    \centering
    \includegraphics[width=1.0\textwidth]{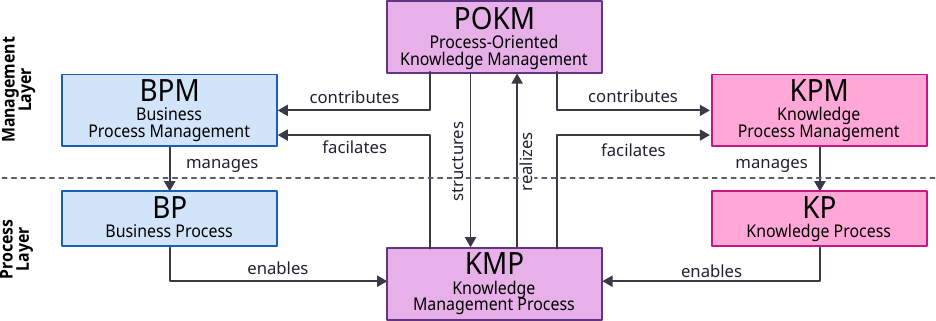}
    \caption{Management Layer and Process Layer: Elements and Their Relationship}
    \label{fig:relations}
\end{figure}

To implement this integration effectively, organizations rely on KMP, which focus on structuring and executing activities related to knowledge creation, storage, retrieval, and dissemination. 
\nk{Generally, considered disciplines can be separated into two conceptual layers: the management layer and the process layer. The management layer is responsible for governance and coordination, while the process layer is responsible for the actual realization (i.e., execution) of business- and knowledge-related activities. The relationships between and within these layers and their elements are illustrated in Figure~\ref{fig:relations}.}

KMP defines knowledge interactions and knowledge flows in knowledge-intensive tasks  among \textit{resources} in a process-oriented manner~\cite{PROMOTE}. The quality of processes heavily depends on the availability and quality of resources. In turn, the quality of resources is influenced by the application of knowledge operational skills in executing activities within a process~\cite{DBLP:journals/jkm/KalpicB06}.

As the effectiveness of KMP relies on the quality and availability of resources, it becomes essential to examine specific elements that enable these processes (i.e., the model, activities, and resources) and their interactions. These interactions should not be analyzed solely at the levels of control flow or data flow but also at the resource level. Resources—whether human, technological, or informational—are fundamental to the execution and optimization of knowledge management process~\cite{DBLP:conf/caise/RussellAHE05}, as they serve as the bridge between processes, knowledge, and execution.

Interactions associated with the resource perspective were explored and documented in ~\cite{DBLP:conf/caise/RussellAHE05}. However, they were defined in a way that ensures their applicability to 
%the broadest possible range of 
process-aware information systems and do not specifically account for knowledge management scenarios. 

As the nature of knowledge processes is different from the nature of standard business processes, these kinds of processes sometimes require patterns that are not already described in the existing workflow resource patterns, or patterns that should be modified according to knowledge process purposes ( i.e., new patterns and modified patters). Such knowledge patterns mostly arise due to necessity of coordination and orchestration of knowledge workflows between different tasks or inside  one task, where knowledge retrieval, communication, and collaboration must be carefully synchronized or aggregated. Existing workflow resource patterns do not sufficiently cover this level of complexity and integration required for effective knowledge process automation~\cite{DBLP:journals/isem/SarnikarZ08}.

There have been several attempts to determine whether resources carry out or require specific behavioral patterns in their knowledge activities within knowledge management systems
~\cite{10.1145/1645164.1645177,DBLP:journals/isem/SarnikarZ08,Thom2007WorkflowPF,workers}.
However, these studies focused on knowledge-intensive business processes (i.e., a dynamically evolving process involving activities that are difficult to plan in advance, where a set of involved users may be discovered as the process scenario unfolds and collaborative interactions among the users is its major part~\cite{DBLP:journals/jodsn/CiccioM015}) and, therefore, solely on human resources.

\section{Beyond Existing Definitions: Knowledge Processes}
\label{sec:kps}

Existing definitions for knowledge(-intensive) (business) processes (KPs) are informal, and typically center around a lack of structure, unpredictability, collaboration, decision-making, and high level of human involvement. To the best of our knowledge there are no papers differentiating KPs from ``normal'' business processes. It is therefore hard to define methods and systems for supporting KPs~\cite{DBLP:journals/bpmj/GoncalvesBSG23}.

Most related work~\cite{DiCiccio2015KnowledgeIntensivePC} assumes that only unstructured processes can be considered KPs, but \cite{DBLP:journals/bpmj/GoncalvesBSG23} assumes KPs also include well-structured, routine processes.
Similar to~\cite{Davenport} we propose to utilize inputs and outputs of KPs to identify and classify KPs. For this we rely on the following definitions for (a) \textbf{process}, (b) \textbf{process elements}, and (c) \textbf{knowledge process assets}, including (d) \textbf{dependencies between knowledge process assets}, which then leads us to the definition of a (e) \textbf{knowledge process categorization} to establish a connection to traditional activity-oriented processes.

\begin{figure}[htb!]
    \centering
    \includegraphics[width=1.0\textwidth]{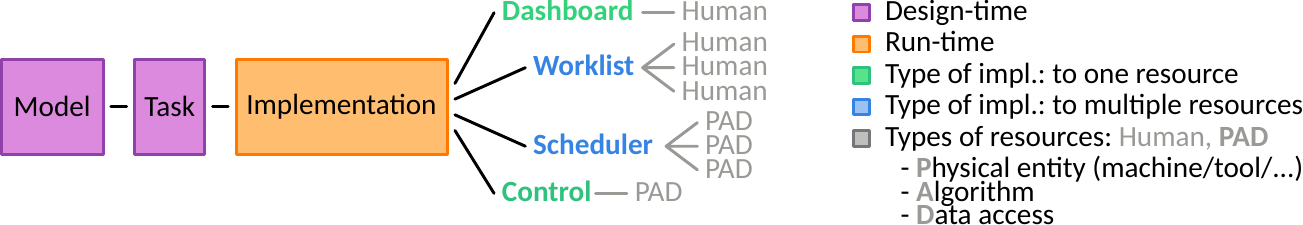}
    \caption{Knowledge Process Elements}
    \label{fig:elem}
\end{figure}

\textbf{(a) Process:} A process is represented by a model, which contains a set of tasks (see Fig. \ref{fig:elem}) that rely on resources for their execution (e.g., a clerk decides if a request is granted or a database has to deliver requested data). The model outlines all tasks that need to be completed in order to execute the process~\cite{DBLP:conf/otm/BrownP05}. The model is abstract, it has to be instantiated for a particular scenario (i.e., a customer, a produced part, or a particular business case). An instantiated model is executed, i.e., all task are performed for a particular scenario. The uncertainty which related work claims to be setting apart knowledge processes from other processes is when relying on this definition a part of every process.

\textbf{(b) Process Elements:} As depicted in Figure~\ref{fig:elem}, a process model contains a set of tasks. Each task is associated with an implementation, the relies on input and generates output. It does so by relaying the input to humans or non-human resources such as algorithms or data access mechanisms. Which particular resource fulfills a request defined through the task depends on the scenario (e.g., a travel expense report has to be handled by different clerks depending if you traveled locally or abroad). Thus the scenario determines which resources should be involved and how these resources perform specific tasks. However, the final set of resources (both human and non-human) can not necessarily be set when the process starts and can potentially only be discovered as the process unfolds~\cite{DBLP:journals/jodsn/CiccioM015}.
Implementations of tasks can vary based on the type of resource (human or non-human) and the number of resources involved (single or multiple) (see Fig.~\ref{fig:elem}).

Decisions regarding tasks (and the model they occur in) are made during the design-time (i.e., all design and analysis activities that occur prior to actual modeling and process execution). During design-time all tasks are parametrized to reflect how resources should be invoked. During run-time tasks are then enacted by concrete resources, and tasks are dynamically assigned and executed by the resources  based on the specifications defined in the model~\cite{time}.

\begin{figure}[htb!]
    \centering
    \includegraphics[width=1.0\textwidth]{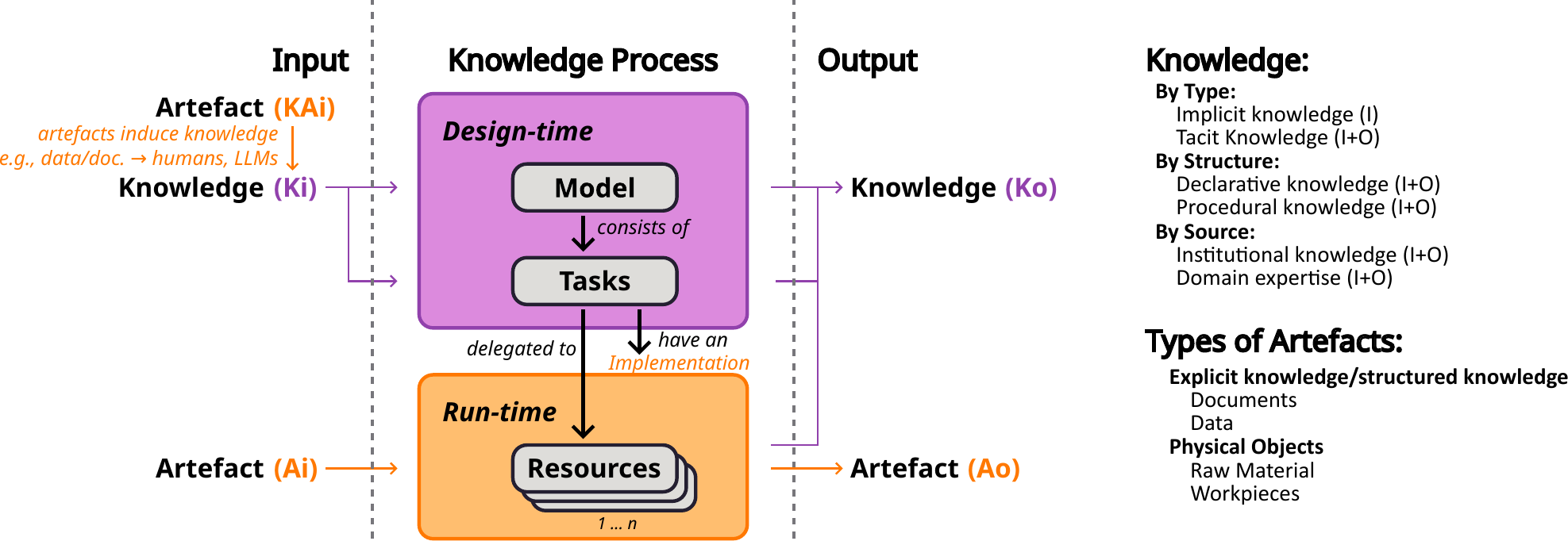}
    \caption{Knowledge Process Assets and Their Dependencies with Knowledge Process Elements}
    \label{fig:bigpic}
\end{figure}

\textbf{(c) Knowledge Process Assets:} For this definition we rely on the concepts depicted in Figure~\ref{fig:bigpic}. Artefacts like data or documents contain knowledge. Artefacts thus contain explicit knowledge, while knowledge itself can be tacit (only contained in the heads of humans). We define \textbf{knowledge process assets} to be \textbf{inputs} required (a) to create a KP (KAi, Ki), (b) to execute a KP (Ai), and \textbf{outputs}, that (c) are produced via transformation during execution (Ao), and (d) are gained after the execution (Ko).

\textit{Input Knowledge} (Ki) is the conceptual, non-physical information that exists in the human heads (mostly process modeler or domain expert), or can be derived out of physical Knowledge artefacts (KAi, like documentation, specification, system requirements, etc.) and are processed by process modeler. This unstructured conceptual knowledge is then transformed into a structured model. Model in this case is a repository of knowledge that defines how things work, where tasks can be considered as a set of guidelines or steps derived from the model, which contain part of the model’s knowledge to perform specific operations.

\textit{Input Artefacts} (Ai) are physical or tangible materials and items to which knowledge can be applied to alter their state. Examples of such artefacts include raw materials, components that need to be assembled, data that must be analyzed or processed, and documents that serve as reference or instructional materials (e.g., blue prints). Resources are entities assigned to carry out specific operations to transform Ai into output artefacts, utilizing either the knowledge provided within the task, their own expertise and skills, or a combination of both. 

It is important to mention, that even so resources themselves have part of knowledge to perform tasks, we do not consider this knowledge as input knowledge, but rather as a resource characteristic that is closely connected with its role and level of experience and skills (i.e., it exists as part of the resource's ability to perform tasks).
\nk{Another aspect that supports this distinction, is that changing the resource that handles a particular task might affect the quality of this task (i.e., its execution and resulting output), even though the Ki remains unchanged. For instance, different workers or agents could interpret the task differently. Similarly, machines with identical functions could execute operations with varying levels of precision (e.g., two machines with different maintenance schedule).} 

\textit{Output Artefacts} (Ao) represent the transformed state of the initial Ai after applying the Ki (i.e., preforming tasks by resources). Examples of such artefacts can be refined raw materials, finished or semi-finished products, processed data (e.g., reports, structured information), and updated or newly created documents (e.g., revised blueprints, manuals, or instructional materials).

\textit{Knowledge Output} (Ko) is the refined or newly generated knowledge gained after process execution, allowing to change or improve the original model, as well as lessons learned, and best practices. This knowledge can be transformed into KAi and contribute to refining methodologies, updating instructional materials, and embedding new knowledge into systems.

\textbf{(d) Dependencies Between Knowledge Process Assets:} To define the types of KPs, it is essential to first consider the nature and role of different KP assets, as each of the assets plays a different role across design- and run-time: 

\begin{enumerate}[noitemsep,label=\textbf{(\roman*)}]
    \item Ki is always required to design process models and tasks. 
    \item Ai may or may not be required depending on the nature of the process. They are typically involved in the creation of Ao, but not all processes rely on artefact transformation. 
    \item Ko can be derived at all levels of the process (model, task, and resource). Ko reflects intellectual or cognitive value of the process. 
    \item Ao are usually produced after tasks are performed by resources, involving a transformation of either the Ki, Ai, or both. Ao reflects the operational output of the process. 
    \item Internal resource knowledge is not treated as Ki, as it is considered an inherent characteristic of the resource itself. This is relevant for both human and non-human resources.
   \item Internal resource knowledge is not immediately process-specific. However, during process execution via learning, innovation, or feedback this general resource knowledge can be transformed into process-specific Ko.
\end{enumerate}

\textbf{(e) Knowledge Process Categorization:}
The interaction between inputs and outputs provides the foundation for categorizing KPs. In Table~\ref{tab:processtype}, we present possible types of knowledge processes, based on required inputs (Ki, Ai) and resulting outputs (Ko, Ao). This classification serves to link each process type to its knowledge intensity level.

\begin{table}[!ht]
    \centering
    \caption{Types of Knowledge Processes}
    \begin{tabular}{p{2.2cm}|p{8.3cm}|p{1.5cm}}
        Process Type                   &  Description & Knowledge Intensity \\ \hline \hline
        Ki    $\rightarrow$     Ao     &  \textbf{Knowledge-based Artefact Generation} & low \\ 
        ~ &  generation of new data based on knowledge input, parsing data from one format to another & \\
        Ki, Ai $\rightarrow$     Ao     &  \textbf{Knowledge-based Artefact Processing} & low\\
        ~ &  a simple, well-established process that is so routine and standardized that people may not actively focus on improving it.& \\
        Ki    $\rightarrow$     Ko     & \textbf{Knowledge Processing} & medium \\
        ~ &  a knowledge-building or reasoning process where information is processed to create new knowledge & \\
        Ki    $\rightarrow$     Ko, Ao &  \textbf{Knowledge-based Artefact Development} & medium\\
        ~ &  a process where conceptual knowledge not only generates new ideas or models but also leads to physical changes (e.g., prototype or document that reflects that new knowledge) & \\
        Ki, Ai $\rightarrow$     Ko     & \textbf{Artefact-based Knowledge Extraction} & high\\
        ~ &  a process like training or simulation, where physical resources interact with conceptual knowledge to derive new insights, conclusions, or models.&  \\
        Ki, Ai $\rightarrow$     Ko, Ao &  \textbf{Artefact Processing with Knowledge Extraction} & high\\
        ~ &  a full-scale operational process where new knowledge is not only generated but also applied to produce tangible results. \\ \hline
        Ai    $\not \rightarrow$ Ko     &  \textbf{Artefact-based Knowledge Generation} & \\
        Ai    $\not \rightarrow$ Ao     &  \textbf{Artefact Processing} & \\
        Ai    $\not \rightarrow$ Ko, Ao &  \textbf{Artefact-based Knowledge Extraction} & 
    \end{tabular}
    \label{tab:processtype}
\end{table}

Since Ki is always required to design process models and tasks, it can be said that every process involves knowledge and, as a result, can be considered a knowledge-intensive process. However, the level of knowledge intensity may vary.

Knowledge-based Artefact Generation (Ki $\rightarrow$ Ao) is considered to be low knowledge intense since it only uses knwoeldge to produce artefacts (i.e., transform information into another representation without creating additional value), being largely structured, predictable, and repeatable, making it more suited to automation. The same is valid for Knowledge-based Artefact Processing (Ki, Ai $\rightarrow$ Ao), as in the core is also the data transformation utilizing in addition to knowledge also some input data.

Knowledge Processing (Ki $\rightarrow$ Ko) and Knowledge-based Artefact Development (Ki $\rightarrow$ Ko, Ao) are considered to be medium knowledge intense. These types of processes involve not only knowledge utilization and interpretation, but also reasoning and generation of new knowledge. However, despite the complexity of these processes, we consider them as medium, as Ki is already conceptual and abstract, allowing for more straightforward reasoning and knowledge generation (i.e., Ko).

Artefact-based Knowledge Extraction (Ki, Ai $\rightarrow$ Ko) and Artefact Processing with Knowledge Extraction (Ki, Ai $\rightarrow$ Ko, Ao) is considered to be the most complicated processes to realize, since transforming Ai (which are often formalized documents or structured data) into new knowledge (i.e., Ko) requires more complex interpretation and processing to extract meaningful insights.

The \textbf{Knowledge Process Categorization} can now be applied to any business process to determine its knowledge intensity. Conversely we also want to emphasize that by the definition presented in this paper \textbf{any business process is a knowledge process}. The process model itself (design time) represents knowledge, and any execution (run-time) produces an output which represents knowledge\footnote{E.g., a process with a single task with no related input and output, e.g., starting a machine, has an implicit output - a workpiece }.

\nk{The proposed Knowledge Process Categorization is especially relevant in the context of the ongoing transition from traditional BPM towards Agentic BPM (ABPM). The central idea of ABPM is the deployment and execution of autonomous software agents to achieve business process goals~\cite{Hoang}. ABPM focuses on goal-oriented execution, where the desired goal is set while the concrete actions for its achievement remain open and can change during agent run-time. Therefore, agent coordination and autonomous decision-making become central aspects for achieving individual agent goals along with the overall process objectives~\cite{formal}.}

\nk{These changes question the traditional understanding of knowledge-intensive processes as human-centered processes. Thus, the categorization proposed in this paper provides a foundation for \textbf{analyzing knowledge intensity independently of involved resource types} (e.g., human, machine, autonomous agents, etc.), since knowledge intensity of the process is not tied to the type of resource executing this process, but rather to the role of assets in the process itself.}

\section{Knowledge Process Patterns}
\label{sec:patterns}

In this section we want to concentrate on how process asset generation and modification is related to process elements (i.e., models, tasks, implementations, and resources) (see Fig. \ref{fig:bigpic}). We want to identify 
the typical implementations of tasks, the typical fragments in the process model to support process assets generation or modification, and the typical ways to interact with assets (i.e., interact with artefacts and knowledge as part of a process execution).

\begin{figure}[htb]
  \centering
  \includegraphics[width=0.5\textwidth]{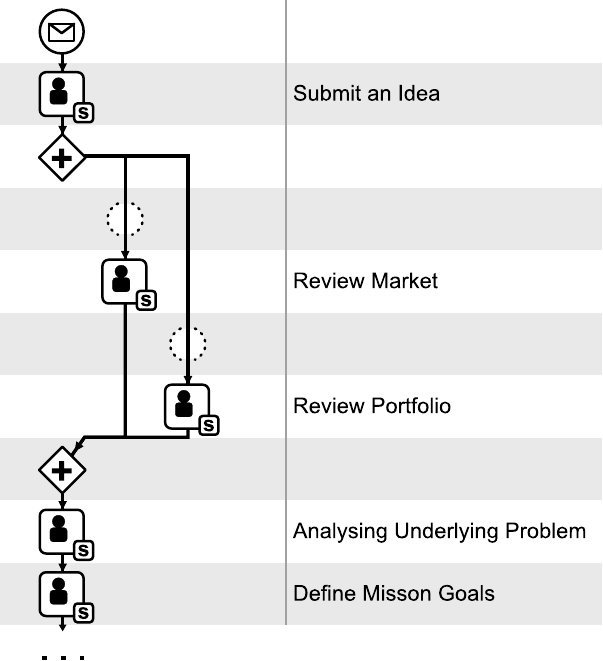}
  \caption{Realization of The Ideation Process}
  \label{fig:ideation}
\end{figure}

Based on the definition of KPs as elaborated throughout Section~\ref{sec:kps} (allowing to categorize KPs), we identify knowledge process assets:  
\begin{itemize}
   \item 
 % i) 
 At design-time, knowledge flows into generating the process model. 
   \item
% ii)
Each task in the process model and its implementation represents the knowledge how to do something.
   \item 
  % iii) 
  The input and output for each task (at run-time) are process assets.
   \item
 % iv) 
 The outputs of tasks lead to new assets (e.g., involved resources are learning about the process and processes are improved).
\end{itemize}

The patterns presented in this section are not related to workflow patterns~\cite{van2003workflow}, as we focus solely on the (a) semantic and implementation of tasks which are performed, (b) the purpose of interactions with assets, and (c) how human resources interact with assets.

In order to derive a list of patterns we utilized a real-world process which facilitates the creation of ideas of a new product or application for a large company (furthermore called IP - ideation process)\footnote{Full version: \url{https://cpee.org/flow/graph.html?load=https://cpee.org/hub/server/Papers.dir/IP.xml}, last access: 20026-06-05}, partially depicted in Fig.~\ref{fig:ideation}.
\nk{The process consists of 13 human tasks, 11 machine tasks, and one subprocess that is repeatedly initiated. It produces a large number of artifacts, as it involves all major types of knowledge processing: acquisition, sharing, application, protection, storage, and creation.} \etm{Although all depicted tasks are modeled and realized as human tasks, the process has been created with the intent to replace human tasks with agent tasks in the future, which will not change the analysis conducted in this section.}

The IP can be categorized as \textbf{Artefact Processing with Knowledge Extraction} (Ki, Ai $\rightarrow$ Ko, Ao, i.e., high knowledge intensity). This process is characterized by high artefact dependency, as each subsequent step relies on input from the previous one. Since multiple employees are involved and depend on shared artefacts, the process also exhibits a high level of communication and collaboration. 

\nk{Based on the interaction between process assets and resources during run-time}, we identified 6 patterns groups:
\begin{itemize}[noitemsep,topsep=0pt]
    \item \textbf{Interaction Patterns:} How are assets created or changed at runtime.
    \item \textbf{Intent Patterns:} Common tasks.
    \item \textbf{Evolution Patterns:} Common process fragments and how assets flow between them.
    \item \textbf{For Human Resources - Asset Communication Patterns:} How asset creation, change, or content is communicated.
    \item \textbf{For Human Resources - Display Patterns:} How process assets are presented to human resources at run-time.
    \item \textbf{Dependency Patterns:} How process assets are connected with each other.
\end{itemize}

Table~\ref{tab:patterns} presents the list of identified patterns. The column \textit{process element} refers to the scope of a particular pattern and the level of its application within the process, i.e., (model, task, or implementation).
Task implies that a pattern is applicable for one particular task, independent of of its implementation.
Model denotes that the pattern spans multiple tasks or sub-processes, i.e., it is represented by a control flow fragment such as a sequence of steps in a loop, or a decision.
Implementation finally describes that the pattern is realized as a component which is invoked during process execution.

{\scriptsize
\begin{longtable}{p{2.2cm}|p{2cm}|p{7.8cm}}
    \caption{Knowledge Process Patterns, * denotes patterns that are closely related to other patters in the list, but were not directly observed in the analyzed process}
     \\
        Pattern & Process Element & Description   \\ \hline
        \multicolumn{3}{c}{\textbf{Asset Interaction Patterns}} \\ \hline
        Create & task & create a new asset\\
        Update/Improve & task & change part of already existing asset \\
        Comment/Review & task & review an existing asset and provide feedback; yields new assets\\
        Merge* & task & combine two or more assets into one \\
        Split* & task & divide one asset into two or more \\
        Freeze & task/impl. & the current version of an asset can no longer be changed \\
        Copy* & task & duplicate an asset \\
        Archive & task/impl. & store an intermediate state of an asset \\
        Delete* & task & remove assets from the run-time context of a process
        \\ \hline
        \multicolumn{3}{c}{\textbf{Asset Intent Patterns}} \\
        \hline
        Approve/Decline & task & accept or decline process assets; yields a new asset; always has consequences. \\
        Single Vote*  & task & same as approve/decline, but new asset can contain more values than yes/no \\
        Multiple Votes & task/model & same as single vote, the results of votes by multiple resources are aggregated \\
        Decision & task & a resource selects between multiple assets \\
        Report & task & present the current state of an asset to a resource \\
        DeLorean* & task & allow to return to a historic version of an asset \\
        Send & task & transfer an asset to an external system   (control over the asset is lost) \\
        Receive* & task & the \textit{Create Asset Interaction Pattern} is applied for a particular purpose \\
        Reply* & task & the \textit{Send Message Pattern} is applied as a response to a previous \textit{Receive}\\ \hline
        \multicolumn{3}{c}{\textbf{Asset Evolution Patterns}} \\
        \hline
        Redo* & model & historic assets are presented to a resource, a new assets are created for the purpose of fixing previous mistakes or getting different results \\
        Refine/Improve & model & a set of assets is changed to serve a particular purpose (e.g., review)\\
        Review & model & historic process assets remain accessible to be validated by resource, e.g., confirming their quality, relevance, or completeness\\ \hline
        \multicolumn{3}{c}{\textbf{For Human Resources: Asset Communication Patterns}} \\
        \hline
        Information & task & support human resources understanding assets without requiring strict actions \\
        Notification* &  task & inform human resources about an asset status (e.g., a change in an asset)  without requiring strict actions; this may indicate that a particular state in the process is reached and further actions might be required \\
        Instruction & task & inform humans resource how to conduct a task, requiring strict actions \\ \hline
        \multicolumn{3}{c}{\textbf{For Human Resources: Display  Patterns}} \\
         \hline
        Assignment & impl. & convey to a resource which other resource is intended to perform a task\\
        Priority & impl. & convey the urgency of a particular task to a resource\\
        Process Progress & impl. & convey the progress for a particular process to a resource, e.g., 1/15 \\
        Process State & impl. & convey the topology of tasks in a process, and the state of tasks of an instance\\
        Task Progress & impl. & convey the current (non-final) state of assets to a resource \\
        \hline
        \multicolumn{3}{c}{\textbf{Asset Dependency Patterns}} \\
        \hline
        Self-contained & model & an asset can be created/used without relying on any other previous assets \\
        Dependent & model & an asset requires one or more assets to exist
    \label{tab:patterns}
\end{longtable}
}

While the analysis of further processes regarding contained patterns might be necessary, in our experience the following list of patterns covers a wide array of use-cases or scenarios. \nk{The presented set of patterns does not constitute a final taxonomy. But it highlights practical concerns that arise when implementing knowledge management processes in operational systems.}

\nk{For example, voting appears to be a simple and well-defined concept. But, usually, even if it is clear which participants are involved, from an implementation perspective (especially in process-oriented solutions) supporting simultaneous voting of multiple participants across different devices can rely on non-trivial logic (e.g., for complex or interactive consensus mechanisms). Depending on the number of participants, dynamic parallel branching is required to explicitly model and realize voting. As a result, such activities are often realized as single tasks (black boxes; outside the process using external tools), and the results are later manually/automatically read by the process. External approaches increase operational overhead and may introduce errors, reduce traceability, lead to information loss, and reduce the understanding of the voting process.}

\nk{Interestingly, although the knowledge input Ki is required for realizing the model, tasks, and corresponding implementation, and the knowledge output Ko can potentially be derived at each of these levels, pure knowledge processing is, in practice, rarely possible in the context of process execution. After each interaction between a resource and Ki, once Ko is produced, some corresponding documentation is typically created as well (e.g., a report, analysis document, spreadsheet, etc.).}

\nk{Furthermore, we observe that the direct transformation of an input artifact Ai into an output artifact Ao occurs primarily during Asset Interaction Patterns. In many other cases, the resource interacts with Ai leads to creation of a completely new output artifact Ao, rather than the transformed version of the original one. For example, based on an invoice, an approval or rejection protocol may be created, while the invoice itself remains unchanged.}

\nk{In addition, during certain other patterns (e.g., Communication or Display Patterns), even when the resource interacts with the Ai without directly creating Ao, on the implementation level, an output artifact Ao is still generated. The resource may not be immediately aware of it, since information about the interaction is logged within the system. In such cases, Ao may take the form of event logs, database entries, or similar system-generated documentation.}

\section{Conclusions}
\label{sec:conc}

In this paper we pursued the quest for more precise definition of knowledge processes at the intersection of knowledge and process management (RQ1). The definition encompasses activity-oriented processes, and how knowledge is represented by them and in them. Classifying processes according to this definition is straight-forward looking at just in- and output. We furthermore dissected a complex and knowledge intensive real-world process which yielded a set of patterns that we deem common in knowledge processes (RQ2).
The shortcomings of this paper are related to RQ2. In order to confirm the list of patterns, more real-world processes need to be dissected, which proved hard as many companies are hesitant to share executable processes. This will have to be tackled in future work, potentially in larger case-study. 
The main \textbf{takeaway} of this work is, that by utilizing the definition, characterization scheme, and patterns presented in this paper, the challenges faced by companies during knowledge process design and implementation, can be partially alleviated, as it (a) becomes possible to identify and name best practices contained in available process models, and (b) ensure that a business process management system provides necessary components to implement a set of knowledge process patterns.

%\bibliographystyle{splncs}
%\bibliography{main}

\begin{thebibliography}{10}

\bibitem{38db798c34c640d8804e7f955f1ce9f3}
Hislop, D., Bosua, R., Helms, R.:
\newblock Knowledge Management in Organizations: a critical introduction. 4th edn.
\newblock Oxford University Press, United Kingdom (May 2018)

\bibitem{doi:10.2307/41165944}
Ruggles, R.:
\newblock The state of the notion: Knowledge management in practice.
\newblock California Management Review \textbf{40}(3) (1998)  80--89

\bibitem{MAHDI2019320}
Mahdi, O.R., Nassar, I.A., Almsafir, M.K.:
\newblock Knowledge management processes and sustainable competitive advantage: An empirical examination in private universities.
\newblock Journal of Business Research \textbf{94} (2019)  320--334

\bibitem{dalkir2013knowledge}
Dalkir, K.:
\newblock Knowledge management in theory and practice.
\newblock routledge (2013)

\bibitem{brocke2010handbook}
Brocke, J., Rosemann, M.:
\newblock Handbook on Business Process Management 1: Introduction, Methods, and Information Systems.
\newblock Int. Handbooks on Information Systems. Springer (2010)

\bibitem{Jordan}
Abuezhayeh, S.:
\newblock Integration between knowledge management and business process management and its impact on the decision making process in the construction sector: a case study of jordan.
\newblock Construction Innovation \textbf{22}(4) (2022)  987--1010

\bibitem{process-knowledge}
Kovačič, A., Vuksic, V., Lončar, A.:
\newblock A process-based approach to knowledge management.
\newblock Economic Research \textbf{19}(2) (01 2006)

\bibitem{al-emran_impact_2018}
Al-Emran, M., Mezhuyev, V., Kamaludin, A., Shaalan, K.:
\newblock The impact of knowledge management processes on information systems: {A} systematic review.
\newblock Int. J. Inf. Manag. \textbf{43} (2018)  173--187

\bibitem{DBLP:journals/jkm/KalpicB06}
Kalpic, B., Bernus, P.:
\newblock Business process modeling through the knowledge management perspective.
\newblock J. Knowl. Manag. \textbf{10}(3) (2006)  40--56

\bibitem{DBLP:journals/jodsn/CiccioM015}
Ciccio, C.D., Marrella, A., Russo, A.:
\newblock Knowledge-intensive processes: Characteristics, requirements and analysis of contemporary approaches.
\newblock J. Data Semant. \textbf{4}(1) (2015)  29--57

\bibitem{DBLP:conf/hicss/MarjanovicF11}
Marjanovic, O., Freeze, R.D.:
\newblock Knowledge intensive business processes: Theoretical foundations and research challenges.
\newblock In: Systems Science. (2011)  1--10

\bibitem{DBLP:journals/smr/KunzleR11}
K{\"{u}}nzle, V., Reichert, M.:
\newblock Philharmonicflows: towards a framework for object-aware process management.
\newblock J. Softw. Maintenance Res. Pract. \textbf{23}(4) (2011)  205--244

\bibitem{DBLP:journals/fgcs/OriolGET21}
Oriol, X., Giacomo, G.D., Esta{\~{n}}ol, M., Teniente, E.:
\newblock Embedding reactive behavior into artifact-centric business process models.
\newblock Future Gener. Comput. Syst. \textbf{117} (2021)  97--110

\bibitem{a9e7410a34844143a44a58bbdaf2b670}
Reijers, H., {Limam Mansar}, S., {Aalst, van der}, W.:
\newblock Product-based workflow design.
\newblock Journal of Management Information Systems \textbf{20}(3) (2003)  229--262

\bibitem{DBLP:conf/modellierung/Breitmayer0PR24}
Breitmayer, M., Arnold, L., Pejic, M., Reichert, M.:
\newblock Transforming object-centric process models into {BPMN} 2.0 models in the philharmonicflows framework.
\newblock In: Modellierung. (2024)  83--98

\bibitem{10.1108/JKM-02-2025-0252}
Sun, Y., Xu, C.:
\newblock Hybrid cognitive authority and algorithmic subjectivity: rethinking knowledge management in ai-driven communication.
\newblock Journal of Knowledge Management (10 2025)  1--26

\bibitem{DBLP:journals/jkm/BakerBTD97}
Baker, M., Barker, M., Thorne, J., Dutnell, M.:
\newblock Leveraging human capital.
\newblock J. Knowl. Manag. \textbf{1}(1) (1997)

\bibitem{AGUILARSAVEN2004129}
Aguilar-Savén, R.S.:
\newblock Business process modelling: Review and framework.
\newblock Int. Journal of Production Economics \textbf{90}(2) (2004)  129--149

\bibitem{yousefiyan2009combination}
Yousefiyan, M.H., Sepehri, M.M.:
\newblock Combination of process and knowledge management.
\newblock Knowledge Creation Diffusion Utilization (2009)

\bibitem{PASCHEK2018182}
Paschek, D., Ivascu, L., Draghici, A.:
\newblock Knowledge management – the foundation for a successful business process management.
\newblock Procedia - Social and Behavioral Sciences \textbf{238} (2018)  182--191

\bibitem{remus2003blueprint}
Remus, U., Schub, S.:
\newblock A blueprint for the implementation of process-oriented knowledge management.
\newblock Knowledge and Process Management \textbf{10}(4) (2003)  237--253

\bibitem{PROMOTE}
Woitsch, R., Karagiannis, D.:
\newblock Process-oriented knowledge management systems based on km-services: The promote® approach.
\newblock In: Practical Aspects of Knowledge Manag. (2002)  398--412

\bibitem{DBLP:conf/caise/RussellAHE05}
Russell, N., van~der Aalst, W.M.P., ter Hofstede, A.H.M., Edmond, D.:
\newblock Workflow resource patterns: Identification, representation and tool support.
\newblock In: CAiSE. (2005)  216--232

\bibitem{DBLP:journals/isem/SarnikarZ08}
Sarnikar, S., Zhao, J.L.:
\newblock Pattern-based knowledge workflow automation: concepts and issues.
\newblock Inf. Syst. {E} Bus. Manag. \textbf{6}(4) (2008)  385--402

\bibitem{10.1145/1645164.1645177}
Yildiz, U., Guabtni, A., Ngu, A.H.H.:
\newblock Towards scientific workflow patterns.
\newblock In: Workshop on Workflows in Support of Large-Scale Science. (2009)  1--10

\bibitem{Thom2007WorkflowPF}
Thom, L.H., Lochpe, C., Reichert, M.:
\newblock Workflow patterns for business process modeling.
\newblock In: BPMDS'07. (2007)

\bibitem{workers}
Nezafati, N., Razaghi, S., Moradi, H., Shokouhyar, S., Jafari, S.:
\newblock Promoting knowledge sharing performance in a knowledge management system: do knowledge workers’ behavior patterns matter?
\newblock Journal of Information and Knowledge Management Systems \textbf{53}(4) (2023)  637--662

\bibitem{DBLP:journals/bpmj/GoncalvesBSG23}
de~A.~R.~Gon{\c{c}}alves, J.C., Bai{\~{a}}o, F.A., Santoro, F.M., Guizzardi, G.:
\newblock A cognitive {BPM} theory for knowledge-intensive processes.
\newblock Bus. Process. Manag. J. \textbf{29}(2) (2023)  465--488

\bibitem{DiCiccio2015KnowledgeIntensivePC}
Ciccio, C.D., Marrella, A., Russo, A.:
\newblock Knowledge-intensive processes: Characteristics, requirements and analysis of contemporary approaches.
\newblock Journal on Data Semantics \textbf{4} (2015)  29--57

\bibitem{Davenport}
Davenport, T., Jarvenpaa, S., Beers, M.:
\newblock Improving knowledge work processes.
\newblock Sloan Manag. Rev. \textbf{37} (01 1996)

\bibitem{DBLP:conf/otm/BrownP05}
Brown, R., Paik, H.:
\newblock Resource-centric worklist visualisation.
\newblock In: On the Move to Meaningful Internet Systems. (2005)  94--111

\bibitem{time}
Caron, F., Vanthienen, J.:
\newblock An exploratory approach to process lifecycle transitions from a paradigm-based perspective.
\newblock In: Enterprise, Business-Process and Inf. Syst. Modeling. (2011)  178--185

\bibitem{Hoang}
Vu, H., Klievtsova, N., Leopold, H., Rinderle{-}Ma, S., Kampik, T.:
\newblock Agentic business process management: Practitioner perspectives on agent governance in business processes.
\newblock In: Business Process Management: Responsible {BPM} Forum, Process Technology Forum, Educators Forum, Springer (2025)  29--43

\bibitem{formal}
Giacomo, G.D., Kampik, T., Kirchdorfer, L., Montali, M., Weinhuber, C.:
\newblock Formal foundations of agentic business process management (2026)

\bibitem{van2003workflow}
van Der~Aalst, W.M., Ter~Hofstede, A.H., Kiepuszewski, B., Barros, A.P.:
\newblock Workflow patterns.
\newblock Distributed and parallel databases \textbf{14} (2003)  5--51

\end{thebibliography}

\end{document}